\documentclass{iau}
\usepackage{graphicx}

\usepackage{sidecap}
\usepackage{wrapfig}

\title[Star Formation in the Galactic Centre]{Young Stellar Objects close to Sgr~A*}

\author[Behrang Jalali et al.]  
{
B. Jalali$^{1}$, 
F. I. Pelupessy$^{2}$,
A. Eckart$^{1,3}$,
S. Portegies Zwart$^{2}$,
N. Sabha$^{1,3}$, 
A. Borkar$^{3,1}$,
J. Moultaka$^{4,5}$,
K. Mu\v{z}i\'{c}$^{6}$, 
L. Moser$^{1}$
}

\affiliation{
1) I. Physikalisches Institut, Universit\"at zu K\"oln, Z\"ulpicher Str. 77, 50937 K\"oln, Germany
\\
2) Leiden Observatory, Leiden University, PO Box 9513, 2300 RA, Leiden, The Netherlands
\\
3) Max-Planck-Institut f\"ur Radioastronomie, Auf dem H\"ugel 69, 53121 Bonn, Germany 
\\
4) Universit\'e de Toulouse; UPS-OMP; IRAP; Toulouse, France
\\
5) CNRS; IRAP; 14, avenue Edouard Belin, F-31400 Toulouse, France
\\
6) ESO, Alonso de Cordova 3107, Vitacura, Casilla 19, Santiago, 19001, Chile
}

\pubyear{2013}
\volume{303}  
\pagerange{001--002}
\jname{The GC: Feeding and Feedback in a Normal Galactic Nucleus}
\editors{Editors}
\begin{document}

\maketitle

\begin{abstract}
We aim at modelling small groups of young stars such as IRS~13N, 0.1~pc away from Sgr~A*, 
which is suggested to contain a few embedded massive young stellar objects. 
We perform hydrodynamical simulations to follow the evolution of molecular clumps orbiting 
about a $4\times10^6 ~ M_{\odot}$ black hole, to constrain the formation and the physical 
conditions of such groups. 

We find that, the strong compression due to the black hole along the orbital radius vector 
of clumps evolving on highly eccentric orbits causes the clumps densities to increase to 
higher than the tidal density of Sgr A*, and required for star formation. 
This suggests that the tidal compression from the black hole could support star formation. 

Additionally, we speculate that the infrared excess source G2/DSO approaching Sgr~A* on a
highly eccentric orbit could be associated with a dust enshrouded star that may have been
formed recently through the mechanism supported by our models. 
\end{abstract}

\firstsection 

\section{Modelling Approach}
We propose a possible scenario for stellar groups close to Sgr~A*, such as IRS~13N (Eckart et al. 2004 and 
2013; Muzic et al. 2008), by investigating star formation conditions on a clump scale within the 
circumnuclear disk (CND). To reach this goal, we study the contrasts between evolution of isolated and 
orbiting model clumps.

We use smoothed particle hydrodynamics (SPH) method to simulate the evolution of clumps in the AMUSE 
framework (Portegies Zwart et al. 2013). The evolution of our model clump is followed using {\it Fi} 
code (Pelupessy et al. 2004) integrated in AMUSE. The main physics in this study is the self-gravity of gas
and the stellar interaction between newly formed stars.

Our model molecular clumps are isothermal and contain 100 $M_{\odot}$ in $<$~0.2~pc radius, similar to
clumps observed in the CND. We probe two different orbits (eccentricity=0.5 and 0.94) and test two 
temperatures (10 and 50 K).

\section{Results}
\begin{figure}[t]
\includegraphics[width=0.48\textwidth]{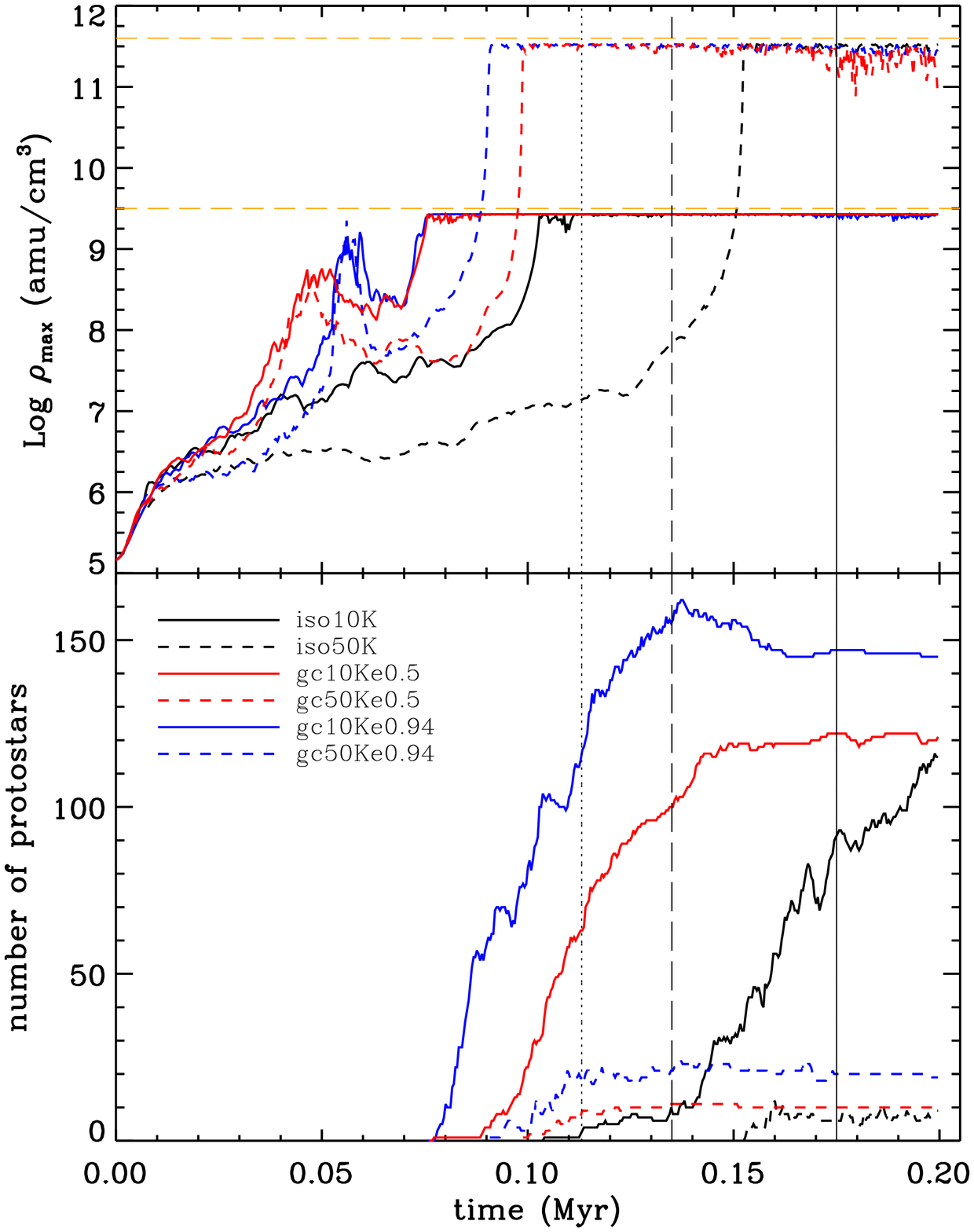}
\includegraphics[width=0.48\textwidth]{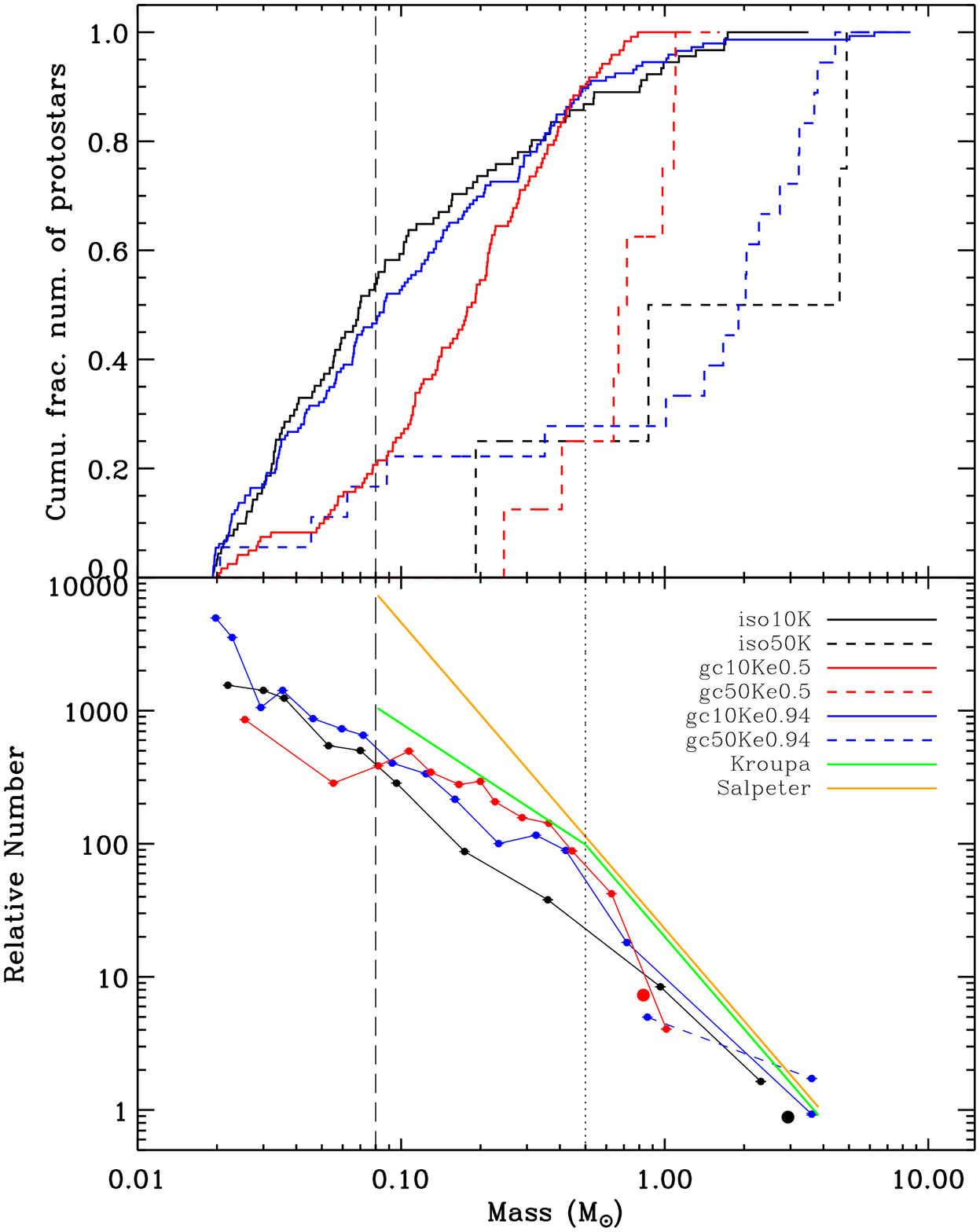}
\vspace*{-.4cm}
\caption{{\bf Left:} The evolution of maximum gas densities and the number of protostars for each model. 
Symbols of all models are consistent in all panels. One orbital period is 0.113~Myr (vertical dotted 
line), one free fall time is 0.135~Myr (vertical dashed line) and the second peri-centre passage is 
0.175~Myr (vertical solid line). Orange horizontal lines are $\rho_{\mathrm{thresh}}$ values for 10 and 
50 Kelvin models, respectively at about 2.69~$\times10^{9}$ and 3.39~$\times10^{11}~\frac{amu}{cm^3}$.
{\bf Right:} The cumulative number of protostars and the IMF of all models computed at 1.3~$\times~t_{ff}\sim$0.175
Myr. In the lower panel, there is only one data point for the iso50K (larger black dot) and gc50Ke0.5 (larger red dot) 
models as there are only a few stars. The Kroupa and Salpeter slopes are overplotted with green and orange lines, 
respectively.}
\label{imf100msun}
\end{figure}

Figure~\ref{imf100msun} shows that the gas densities in the highly eccentric models reach the 
threshold densities (above the SMBH's tidal density) earlier than other models. Consequently, 
protostars (modeled using sinks) form when the threshold densities are reached. The IMF of stars in all 
models are consistent with Kroupa 2001 slope in 0.1-0.5 ~$M_{\odot}$ range.

\begin{wrapfigure}{R}{0.45\textwidth}
\vspace*{-.5cm}
\begin{center}
\includegraphics[width=0.45\textwidth]{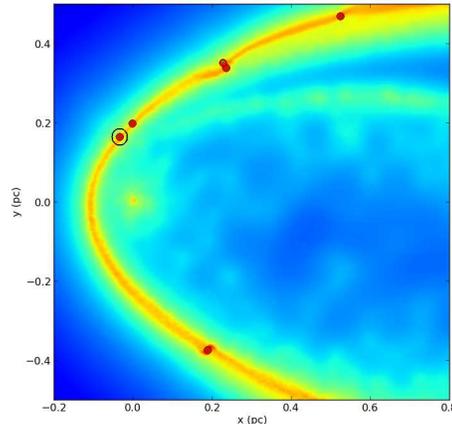}
\end{center}
\vspace*{-.3cm}
\caption{The column density of 50 K model with e$\sim$0.95 at the peri-center passage. The gas density 
increases from blue to red on logarithmic scale. The black hole is situated at the origin (0,0).}
\label{rhomap}
\vspace*{-1.3cm}
\end{wrapfigure}
The gas column density of the highly eccentric 50 K model after 0.17 Myr is shown in Figure~\ref{rhomap}. 
The black circle marks a group of stars with properties similar to that of IRS 13N. Note that individual stars 
are indistinguishable in this image due to the large physical scale and projection effects.

\section{Conclusions}
Our results show that tidal compression from the black hole can support star formation in its vicinity. In the
highly eccentric model clumps, a few groups of young stars with properties similar to IRS 13N are formed  
(Table 4 in Jalali et al; submitted). Our current models confirm the episodic star formation in the 
Galactic Center region. The object G2/DSO (Gillessen et al. 2012, Eckart et al. 2013) could have originated from
a young star that has formed in a clump similar to our highly eccentric model clumps close to the SMBH.


\end{document}